\documentclass[a4paper]{JHEP3} 
\usepackage{microtype}
\usepackage{lmodern}
\usepackage[T1]{fontenc}
\usepackage{amssymb}
\usepackage{color}
\let\normalcolor\relax 
\definecolor{myred}{rgb}{0.6,0,0} 
\definecolor{myblue}{rgb}{0,0.2,0.4}
\definecolor{mygreen}{rgb}{0,0.9,0.1}
\definecolor{hc}{rgb}{.9,0.1,0.7}
\definecolor{hcout}{rgb}{.9,0.7,0.9}
\definecolor{Orange}{rgb}{1.,0.65,0.}
\usepackage{amsmath}
\usepackage{rotating}
\usepackage{timestamp} 

\numberwithin{equation}{section}
\numberwithin{figure}{section}
\numberwithin{table}{section}

\newcommand{\be}{\begin{equation}}
\newcommand{\ee}{\end{equation}}
\newcommand{\bea}{\begin{eqnarray}}
\newcommand{\eea}{\end{eqnarray}}

\title{%
Left-right symmetry at LHC \\and 
  precise 1-loop low energy data 
}

  \author{J. Chakrabortty~${}^{a}$,
  J. Gluza~${}^{b}$,
  R. Sevillano~${}^{b}$,
  R. Szafron~${}^{b}$,
  
  \\
  $^{a}$~ Theoretical Physics Department, 
Physical Research Laboratory \\
Navarangpura, Ahmedabad - 380009, India
 \\
  $^{b}$~Department of Field Theory and Particle Physics,\\
    Institute of Physics, 
    University of Silesia,\\  Uniwersytecka 4, PL-40-007 Katowice,
    Poland

\\

\email{joydeep@prl.res.in},
\email{gluza@us.edu.pl},
\email{robert.szafron@us.edu.pl},
\email{radomir.sevillano-borkowski@us.edu.pl}
}

\abstract{
Despite  many tests, even the Minimal Manifest Left-Right Symmetric Model (MLRSM) has never been ultimately confirmed or falsified. LHC gives a new possibility to test  directly the most conservative version of left-right symmetric models  at so far not reachable energy scales. 
If we  take into account precise limits on the model which come from low energy processes, like the muon decay,
possible LHC signals are strongly limited through the correlations of parameters among heavy neutrinos, heavy gauge bosons and heavy Higgs particles. 
To illustrate the situation in the context of LHC, we consider  the "golden" process $pp  \to e^+ N$.  
For instance, in a case of degenerate heavy neutrinos and heavy Higgs masses at 15 TeV (in agreement with FCNC bounds) we get 
$\sigma(pp  \to e^+ N)>10$ fb at $\sqrt{s}=14$ TeV which is consistent with muon decay data for a very limited $W_2$ masses in the range (3008 {\rm GeV}, 3040 {\rm GeV}).
Without restrictions coming from the muon data, $W_2$ masses would be in the range (1.0 TeV, 3.5 TeV). 
Influence of heavy Higgs particles themselves on the considered LHC process is negligible (the same is true for the light, SM neutral Higgs scalar analog). 
In the paper decay modes of the right-handed heavy gauge bosons and heavy neutrinos are also discussed. Both scenarios with typical see-saw light-heavy neutrino mixings and the mixings which are independent of heavy neutrino masses are considered. In the second case heavy neutrino decays to the heavy charged gauge bosons not necessarily dominate over decay modes which include only light, SM-like particles. 
}


\preprint{LPN- 12-044}

\begin{document}

\section{Introduction}

In general there are two ways in which non-standard models can be tested. In the first approach, Standard Model (discovered) processes or observables can be calculated very accurately by taking into account
 radiative corrections of the non-standard model. 
In the second approach we can look into completely new effects (new processes) which are not present in the Standard Model (SM) but exist in its extensions.  Their detections would be a clear signal for the non-standard physics. Here radiative corrections beyond leading order are, at least at first approximation, not necessary.

At the LHC era it is interesting to think closer how these two approaches could be joined and how we can profit from this situation. It is not a common strategy, especially as Grand Unified Theories (GUT) are concerned. Here we calculate 1-loop radiative corrections at low energies consistently in the framework of the non-standard model (not only in its SM subset, this issue of consistency has been explored intensively in  \cite{Czakon:2002wm,npb,epj}, see also \cite{Chen:2005jx,Chankowski:2006hs,Chankowski:2006jk}). In the next step we are looking into some specific non-standard process at LHC, taking into account obtained earlier precise low energy predictions for parameters of the model.

We consider left-right symmetric model based on the $SU(2)_L \otimes SU(2)_R \otimes U(1)_{B-L}$ gauge
group \cite{pati2,georgi} in its most restricted form, so-called Minimal Left-Right Symmetric
Model ($MLRSM$).  We choose to explore  the most popular version of the model
with a Higgs representation with a bidoublet $\Phi$ and two (left and right) triplets $\Delta_{L,R}$ \cite{class}.  We also assume that the
vacuum expectation value of the left-handed triplet $\Delta_{L}$ vanishes, $\langle\Delta_{L}\rangle =0$ and the CP symmetry can be violated  by
complex phases in the quark and lepton mixing matrices. Left and right
gauge couplings are chosen to be equal, $g_L=g_R$. 
For reasons discussed in \cite{Czakon:2002wm} and more extensively in \cite{tev}, we discuss see-saw  diagonal light-heavy neutrino mixings. 
It means that $W_1$ couples mainly to light
neutrinos, while $W_2$ couples to the heavy ones. 
$Z_1$ and $Z_2$ turn out to couple to both of them \cite{gl92,ann}. $W_L-W_R$ mixing is neglected here\footnote{As an interesting detail, the most stringent data comes from astrophysics through the supernova explosion analysis \cite{barbsn1987,lang}, $\xi <3\cdot 10^{-5}$, typically $\xi<0.05$ \cite{pdg2011}.}.

Taking such a restricted model, easier its pa\-ra\-me\-tri\-za\-tion and less extra parameters are involved in phenomenological studies. However, it does not mean that it is easier to confirm or falsify it, in fact,  despite of many interesting studies and constraints, the model has not been ruled out so far (though many interesting questions and problems calling for consistency of the model have been arose \cite{Czakon:2002wm,npb,epj}. 
PDG \cite{pdg2011} gives $M_{W_2}>1$ TeV for standard couplings decaying to $e \nu$,
recently the CMS collaboration established the generic bound \cite{cms} 
$M_{W_2}>1.4$\; {\rm TeV}.
Moreover, CMS published exclusion limits for LR model \cite{cms2}, they excluded large region in parameter space ($M_N$, $M_{W_2}$) which extends up to $M_{W_2}=1.7$ TeV.
Similarly, ATLAS collaboration   gives exclusion  limits on both $M_N$ and $M_{W_2}$. They obtained that $M_{W_2}> 1.8$ TeV for difference in mass of $M_N$ and $M_{W_2}$ larger than $0.3$ TeV \cite{atlas} (for 34 $pb^{-1}$).
The very last ATLAS analysis \cite{atlaslast} based on the integrated luminosity of 2.1 $fb^{-1}$ pushed it even further, for some neutrino mass ranges it reaches already 2.3 TeV.
These exclusion searches assume generally that  $M_{W_2}> M_N$, however in LR model  the situation can be different i.e. $M_{W_2} < M_N$.
Let us note that $K_L-K_S$ data gives for the minimal LR model a strong theoretical limit, which is (at least) at the level of 2.5 TeV \cite{klks1,Maiezza:2010ic}.

In further studies we take then the rough $K_L-K_S$ limit for $W_2$ mass (to which the LHC analysis approaches quickly, and rather sooner than later will overcome it) 
  \begin{equation}
M_{W_2}>2.5\; {\rm TeV}.
\label{cms1}
\end{equation}
  \\
For heavy neutrino limit $M_N>780$ GeV \cite{atlas}, but it  must be kept in mind that bounds on $M_N$ and $M_{W_2}$ are not independent from each other. 
Let us mention that simultaneous fit to low energy charge and neutral currents give $M_{W_2}>715$ GeV \cite{pdg2011,plb}. 
  
Neutrinoless double beta decay allows for heavy neutrinos with relatively light masses, if Eq.(\ref{cms1}) holds, for more detailed studies, see e.g. \cite{Maiezza:2010ic,Chakrabortty:2012mh}.
 
Detailed studies which take into account potential signals with $\sqrt{s}=14$ TeV at LHC conclude that heavy gauge bosons and neutrinos 
can be found with up to 4 and 1 TeV, respectively,  for typical LR scenarios  \cite{Nemevsek:2011hz,Bansal:2009jx,Ferrari:2000sp}.  
Anyway, such a relatively low (TeV) scale of the heavy sector is theoretically possible, even if GUT gauge unification is demanded, for a discussion, see e.g. \cite{Shaban:1992vv} and \cite{Lindner:1996tf}.

As far as one loop corrections are concerned, there are not many papers devoted to the LR model.  
Apart from \cite{Czakon:2002wm,npb,epj,ann} in which one of the authors of this paper has been involved (MLRSM model), there are other papers: \cite{beal} (limits on $W_2$ mass coming from the $K_L-K_S$ process (finite box diagrams, renormalization not required), 
\cite{pil} (LEP physics), \cite{bsg} (process $b\to s \gamma$). Some interesting results are included also in papers \cite{sok} where the problem of decoupling of heavy scalar particles in low energy processes has been discussed.

On the other hand, the LHC collider gives us a new opportunity to investigate LR models and to look for possible direct  signals. 
  Lately a few interesting papers analysed possible signals connected with the LR model \cite{Nemevsek:2011hz,Ferrari:2000sp,Maiezza:2010ic,Tello:2010am,Frank:2011rb,Jezo:2012rm}. As we are looking for non-standard signals, we restrict here calculations at high energies to the first approximation (tree level).
  
In the next section we will discuss low energy limits on right sector of MLRSM which come from precise calculation of the muon decay.
In section \ref{lhcex} some representative LR signals at LHC will be discussed, taking into account severe limits coming from the muon decay analysis.
We end up with conclusions. We have decided to skip most of the details connected with definition of fields, interactions and parameters in the MLRSM. All these details can be found in \cite{ann} and \cite{npb} (especially the Appendix there). 
  
\section{One-loop low energy constraints on the right sector in MLRSM \label{seclow}}

Four-fermion interactions describe low energy processes in the limit $\frac{q^2}{M^2} << 1$, where $q$ is the transfer of four momentum  and $M$ is the mass of the gauge boson involved in the interactions.  This is an effective approximation of the fundamental gauge theory. This construction allows to replace the complete interaction by the point interaction with the {\it effective} coupling constant (which depends on the model). Independently, the model can be postulated with universal constant coupling  (e.g. Fermi model with {\it universal} constant $G_F$).
Next, taking into account the perturbation, corrections to so defined constants can be calculated at higher levels. Both {\it effective} and  {\it universal} procedures 
describe the same process, so the corrections calculated in this way  must be the same. This fact can be used to constrain parameters of the tested model.

In the SM,  all radiative corrections are embedded in the $\Delta r$ term \cite{pdg2011}
\begin{equation}
\frac{G_F}{\sqrt{2}}=\frac{e^2}{8(1-M_W^2/M_Z^2)M_W^2}(1+\Delta r).
\label{deltar}
\end{equation}
\\
With the present values of the coupling constants and masses \cite{pdg2011}
\begin{eqnarray}
\label{val}
 && G_F = 1.166364(5)\cdot 10^{-5}\;{\rm GeV}^{-2}, \;\;\;\;
  1/\alpha = 137.0359976 \pm 0.00000050, \nonumber \\
  && M_{W} =  80.399 \pm 0.023 \; {\rm GeV},\;\;\;\;
  M_{Z} =  91.1876 \pm 0.0021 \; {\rm GeV},
\end{eqnarray}
\\
experimental fits to the $\Delta r$ parameter in SM give \cite{pdg2011}\footnote{The error has decreased about 3 times during last decade or so, mostly due to improvements in $W$ boson mass measurement.}
\begin{equation}
\label{dr}
\Delta r \equiv \Delta r_0 \pm \Delta r_{\sigma}= 0.0362 \pm 0.0006.
\end{equation}

Matching for the muon decay and the structure of $ \Delta r$ in the  MLRSM model at the 1-loop level has been discussed in \cite{npb}, see also \cite{matching} for more details on the matching in the context of SM.

\FIGURE{
\includegraphics[width=.6\textwidth]{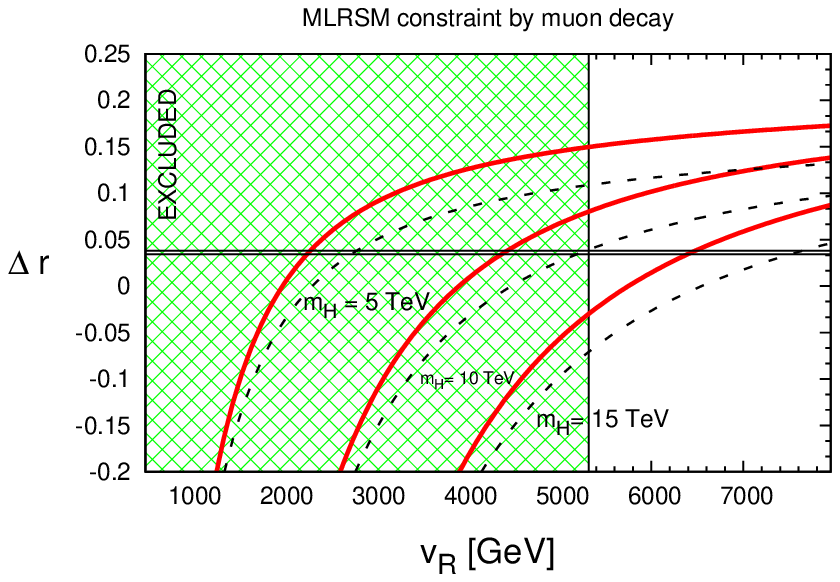}
\caption[]{$\Delta r$ as function of ${\rm v}_R$ for different masses of heavy Higgs particles  5 TeV,  10 TeV and 15 TeV, see Eqs.(\ref{ha},\ref{hb}). Solid (dashed) lines are for neutrino heavy masses with $h_M=0.1$ $(h_M=1)$, see Eq.(\ref{hm}). 
Bold horizontal lines show the  3$\sigma$ C.L. constraint on $\Delta r$, see Eq.(\ref{dr}). Excluded region comes from the  bound  on $W_2$, Eq.~(\ref{cms1}).}
\label{muon1}
}

In Fig.\ref{muon1} $\Delta r$ as function of ${\rm v}_R$ for different masses of heavy Higgs particles and heavy neutrinos is shown.
While plotting we have considered the variations of $\Delta r$ with respect to ${\rm v}_R$, as the heavy gauge boson masses are directly proportional to this parameter,
\begin{equation} 
M_{W_2} \simeq 0.47 \;{\rm v}_R,\;\;\; M_{Z_2}\simeq 0.78 \; {\rm v}_R,
\label{mw2mz2}
\end{equation}
 see Fig.5 in \cite{npb}.
 Mass of the lightest neutral Higgs scalar is assumed to be $M_{H_0^0}=120$ GeV ($\Delta r$
is not sensitive to this mass, see Fig.6 in \cite{npb}). Masses of remaining heavy Higgs particles
$
H \equiv \{H_1^0,H_2^0,H_3^0,A_1^0,A_2^0,H_1^+,H_2^+,\delta_L^{++}\equiv H_1^{++},\delta_R^{++}\equiv H_2^{++} \}
\label{higgsmass}
$

\begin{eqnarray}
  M_{H} & \equiv & M_{H_1^0}=M_{H_3^0}=M_{A_1^0}=M_{A_2^0}
  =M_{H_1^+}=M_{H_2^+}= M_{H_1}^{++}= M_{H_2^0}=M_{H_2}^{++}  \label{ha} 
\end{eqnarray}
are assumed to be equal, 
\be m_H ={\rm v}_R. \label{hb}
\ee

 Heavy neutrino masses 
\begin{equation}
M_N = \sqrt{2} h_M {\rm v}_R,
\label{hm}
\end{equation}
are taken in the range $h_M \in [0.1,1]$.  $h_M$ is the Yukawa coupling connected with the right-handed Higgs triplet. 
$h_M<0.1$ are not forbidden, however attention should be paid to the limits coming from direct experimental searches (LEP $Z_1$ decays, ATLAS, CMS), especially for a region of small ${\rm v}_R$ which we explore. 
On the other hand, $h_M>1$ reaches non-perturbative region.   
 
We can see, as expected in the framework of GUT models to which MLRSM belongs, that for given $m_H$ and $M_N$ there is a very narrow space for ${\rm v}_R$ which are consistent with muon data (fine-tuning).

\TABLE{
\begin{tabular}{|c|c|c|c|c|}
\hline \hline
{\bf set A} & $m_H=4$ TeV & $m_H=5$ TeV & $m_H=10$ TeV & $m_H=15$ TeV \\
  \hline  
${\rm v}_R$ [GeV]& $(1809,2263) $ & $(2257,2795)$ & $(4373,5283)$ & $(6398^\ast,7639)$ \\  
$M_N$ [GeV] & $ (256,3200) $ & $ (319,3952) $ & $(618,7471) $ & $(905,10803) $ \\
\hline
\hline
{\bf set B}  & $m_H=4$ TeV & $m_H=5$ TeV & $m_H=10$ TeV & $m_H=15$ TeV \\
             & ${\rm v}_R=2214$ GeV & ${\rm v}_R=2738$ GeV & ${\rm v}_R=5189$ GeV & ${\rm v}_R=7513$ GeV    \\
$M_N$ [GeV] & $ (2974,3131) $ & $ (3717,3872) $ & $(6974,7336) $ & $(10118,10623) $ \\  
\hline \hline
\end{tabular}
\caption[]{{\bf Set A. }Values of ${\rm v}_R$ for  which various Higgs masses   give $\Delta r$  in agreement with Eq.(\ref{dr}). 
The ranges of $v_R$  are achieved by varying heavy neutrino masses in the domain $h_M \in [0.1, 1]$, see Eq.(\ref{hm}).
For ${({\rm v}_R)}_{min}$ depicted by an asterisk in the last column, and corresponding $(M_N)_{min}$ for which $h_M<0.1$, see the main text. 
\newline {\bf Set B.} ${\rm v}_R$ is fixed in addition, leaving as the only free MLRSM parameter the neutrino 
mass $M_N$, see Fig.\ref{scheme}. Values obtained for $m_H \leq 5$ TeV  do not fulfil direct LHC experimental search limits, the same is true for $m_H=10$ TeV if the limit Eq.(\ref{cms1}) is applied.
} 
\label{tabrange1}
}

\FIGURE{
{\includegraphics[width=.6\textwidth]{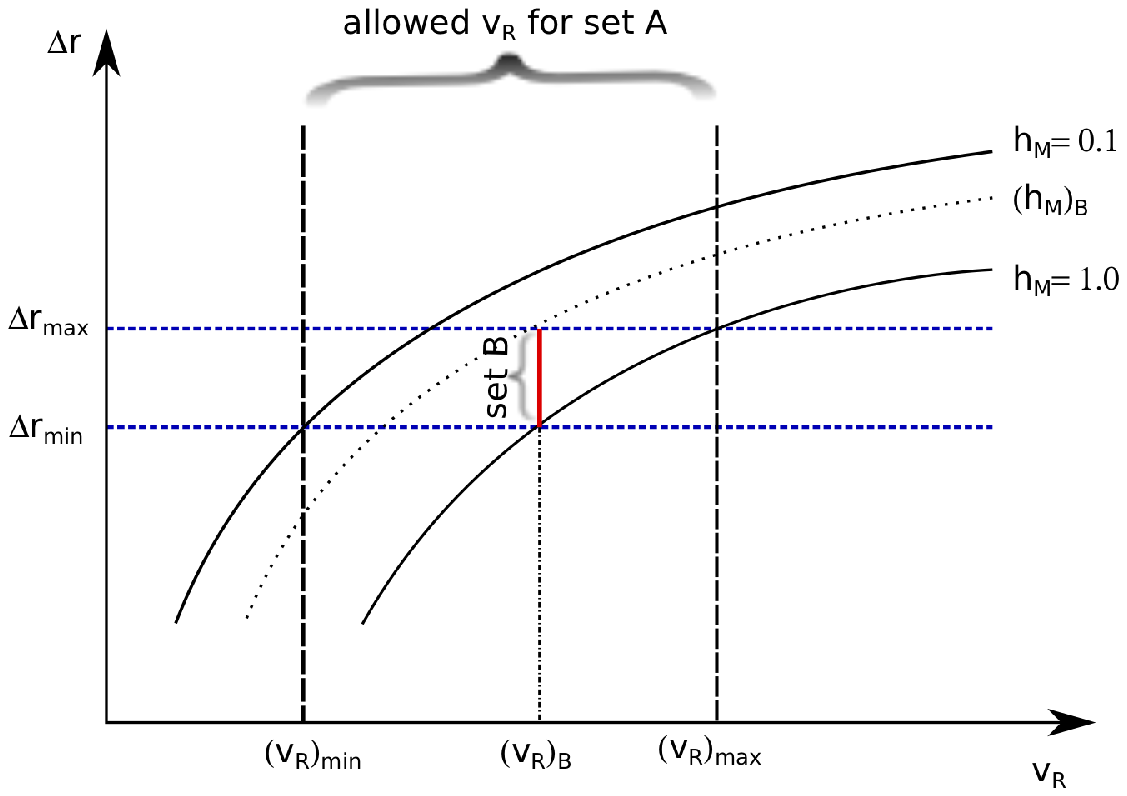}}
\caption[]{{Scheme for  limited parameters in Table~\ref{tabrange1}. ${({\rm v}_R)}_B$ is a fixed value of ${\rm v}_R$ for which set B is defined with maximal value of degenerate heavy neutrino mass (in the perturbative region, $h_M=1$).}} 
\label{scheme}
} 

\FIGURE{
\centering
{\includegraphics[width=.6\textwidth]{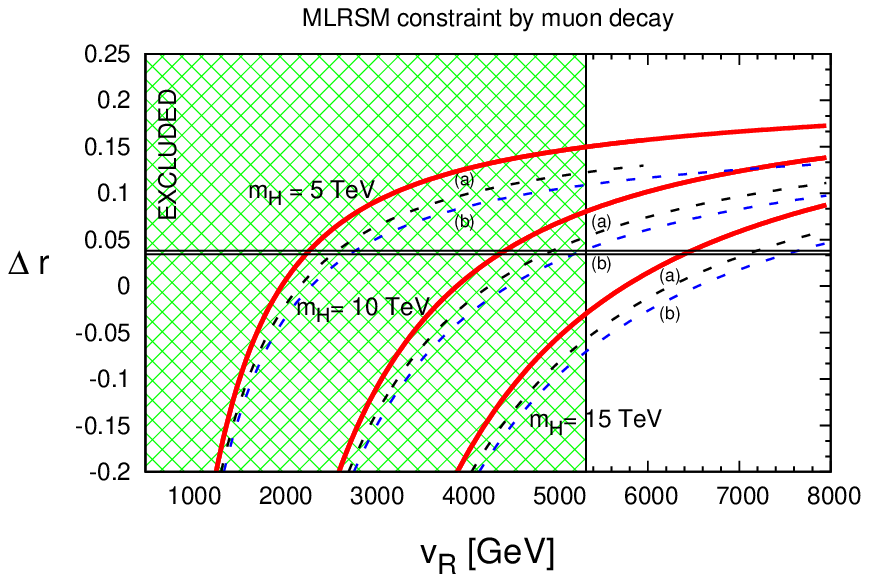}}
\caption[]{$\Delta r$ as function of ${\rm v}_R$ for three different masses of heavy Higgs particles, 5 TeV, 10 TeV and 15 TeV. 
Cases (a) and (b) are different by the heavy neutrino mass spectrum. In the case (a) a mass of $N_4$ is fixed, $M_{N_4}=800$ GeV and masses of $N_5,N_6$ neutrinos vary with ${\rm v}_R$, Eq.(\ref{hm}).
In the case (b)  all  $N_4,N_5,N_6$ neutrinos have masses which obey Eq.(\ref{hm}).
For the  solid lines with $h_M=0.1$ the neutrino cases (a) and (b)  give the same predictions.  
Bold horizontal lines show the  3$\sigma$ C.L. constraint on $\Delta r$, see Eq.(\ref{dr}).}
\label{muon2}
}

Table~\ref{tabrange1} describes the situation more precisely. {\bf Set A} shows ranges of ${\rm v}_R$ which fit at 3$\sigma$ C.L. to Eq.(\ref{dr})
for varying heavy neutrino masses in the range $h_M \in [0.1, 1]$, see Eq.(\ref{hm}). The upper limit of ${\rm v}_R$ corresponds to neutrino masses with $h_M=1$ and $\Delta r_{max}=\Delta r_0
+3 \Delta r_{\sigma}$, the lower limit of ${\rm v}_R$ corresponds to $h_M=0.1$ and $\Delta r_{min}=\Delta r_0
-3 \Delta r_{\sigma}$, see Fig.\ref{scheme}. We can see that the heavy degenerate neutrinos can be relatively light having masses below 1 TeV.
A minimal heavy neutrino mass for $({\rm v}_R)_{min}$ depicted with asterisk in the last column could be even smaller (if $h_M<0.1$).
For instance, ${\rm v}_R=6398$ GeV ($M_{W_2} \simeq 3$ TeV) and muon data in the range 
$\Delta r_0 \pm 3 \Delta r_{\sigma}$ restricts allowed
heavy neutrino masses to the region $100 \leq M_N \leq 2210$ [in GeV] (it means that $(h_M)_{min}\simeq 0.01$). 
{\bf Set B} describes a range of $M_N$ which fits at 3$\sigma$ C.L. to Eq.(\ref{dr})
where in addition also ${\rm v}_R$ is fixed. 
Here a fixed point is chosen to be a value of ${\rm v}_R$ which for given $m_H$ and a neutrino mass with $h_M=1$ gives
$\Delta r_{min}=\Delta r_0 -3 \Delta r_{\sigma}$ (crossing with lower of horizontal lines in Fig.\ref{muon1}). 
Then we are looking for $h_M<1$ which still covers $3 \sigma$ C.L. region constraint by  Eq.(\ref{dr}) and we get the range of neutrino masses written in the Table~\ref{tabrange1}, see Fig.\ref{scheme}.

For {\bf Set B}  possible values of $M_N$ are of course even more restricted than for {\bf Set A}.

Results in Table~\ref{tabrange1} are compatible with  Eq.(\ref{cms1}) for the last column, $m_H=15$ TeV.
 If we take into account FCNC, neutral heavy Higgs mass should be larger than 10-15 TeV, 
going down to a few TeV only in some special cases  (for references and update discussion, see \cite{FCNC}).
So, from now on, let us focus on the last column, $m_H=15$ TeV.
 If we start with some other value of ${\rm v}_R$ instead $({\rm v}_R)_{B}$, e.g. ${\rm v}_R=6500$ GeV ($M_{W_2} \simeq 3055$ GeV) and muon data in the range 
$\Delta r_0 \pm 3 \Delta r_{\sigma}$ restricts allowed
heavy neutrino masses to the region $2654 \leq M_N \leq 3232$ GeV. 

To discuss a case with non-degenerate neutrinos,  
in Fig.\ref{muon2} we  let one of the heavy neutrinos to be much lighter, $M_{N_4}=800$ GeV (for $N_5,N_6$ we keep masses through the relation Eq.(\ref{hm})).
We call it the case (a). For the case (b) we vary all three heavy masses with ${\rm v}_R$, in accordance with Eq.(\ref{hm}) (degeneracy, the same $h_M$). 
In the case $h_M=0.1$ there is only one line, as two cases (a) and (b) give the same predictions. We can see that lines change slightly with chosen neutrino mass spectrum, but not dramatically, values of allowed ${\rm v}_R$ are relatively stable and well constrained.

In summary,   heavy ($m_H > 10$ TeV) Higgs masses are allowed and follow roughly ${ \rm v}_R$ scale
(allowed ${\rm v}_R$ increases with increasing $m_H$). 
However, the most important for the LHC phenomenology is the fact that still light (at the level of hundreds of GeV) heavy neutrinos are allowed in the framework of MLRSM. Let us discuss it more carefully.

\section{Consequences of low energy constraints for MLRSM signals at LHC \label{lhcex}}

\subsection{Decay widths and branching ratios of the heavy LR spectrum}

\FIGURE{
{\includegraphics[width=.49\textwidth]{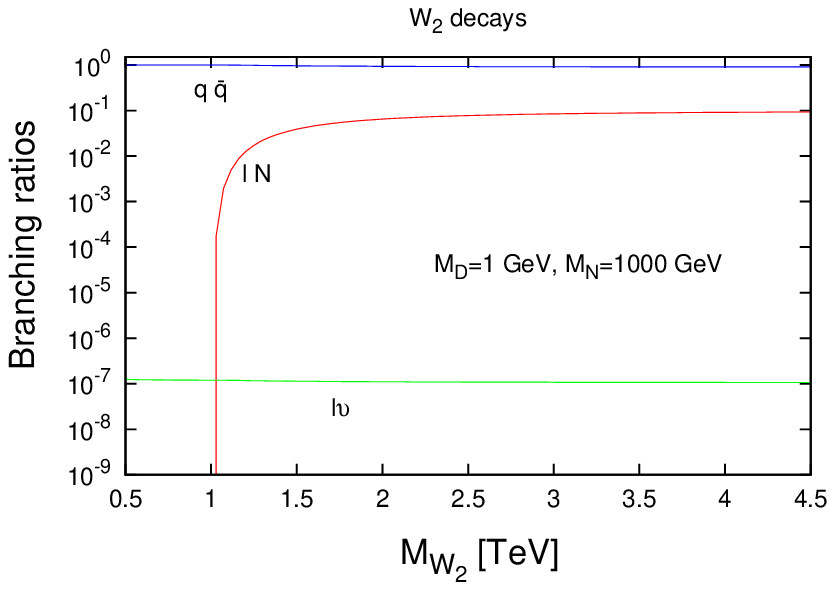}}
{\includegraphics[width=.49\textwidth]{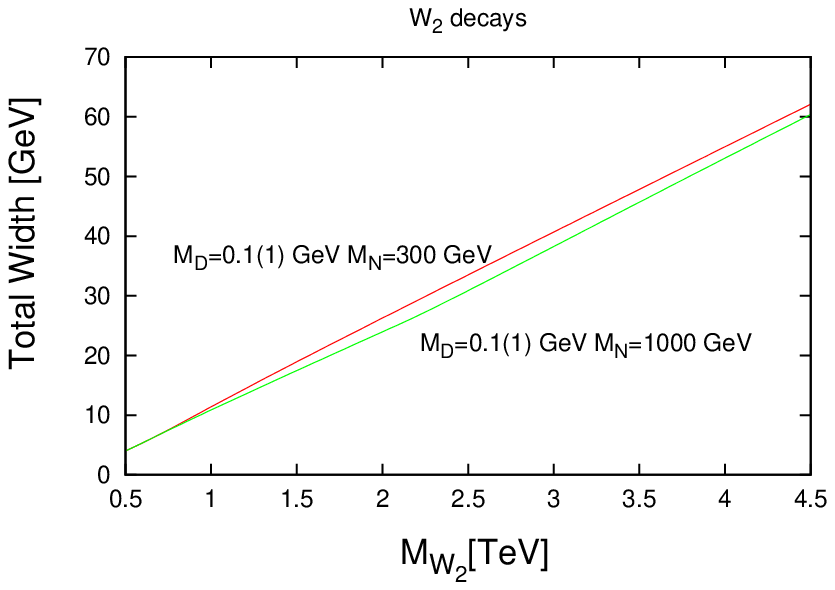}}
\caption{{Decay branching fractions and total widths for $W_2$ decays. Symbol $q \bar{q}$ on this and next plots stands for a sum of all quark flavours,
$q \bar{q}\equiv\sum\limits_{i, i'=u,d,s,b,c,t} q_i \bar{q_{i'}}$. Similarly, $lN \equiv \sum\limits_{i=4}^6 l_{i-3}N_i$, 
$l \nu \equiv \sum\limits_{i=1}^3 l_{i}\nu_i$. }}  
\label{totW2}
}
  
Experimental limits on $W_L-W_R$ mixing angle $\xi$ are very severe and, similarly as in the muon decay case, we neglect it here.
Second,  as already mentioned in Introduction, we assume MLRSM  with diagonal light-heavy neutrino mixings of the 
"see-saw" type

\begin{equation}
|U_{\nu_i j}| \simeq \frac{|\langle M_D \rangle|}{M_{N_j}} \delta_{i,j-3},\;\;\; i=1,2,3,\;\;\;j=4,5,6
\label{mix}
\end{equation}
\\
where $\langle M_D \rangle$ is an order of magnitude of the Dirac neutrino mass matrix and $\nu_i$ stands for 3 light neutrinos.

These two are conservative assumptions, on the other hand they are very natural and we can see what signals we can get at LHC for such  
harsh model conditions. For instance, analyzed in \cite{triplegauge} signals which stem from the gauge boson triple vertices including heavy gauge bosons are absent completely in our scenario.

In Fig.\ref{totW2} we can see that heavy gauge boson decay is dominated by quark channels\footnote{In Fig.\ref{totW2} and the next we  do not depict explicitly exclusion regions  (e.g. Eq.(\ref{cms1})), as the limits for the heavy particle spectrum change quickly with increasing LHC luminosity, see  e.g. \cite{atlas} vs. \cite{atlaslast}.}. Second of importance is $W_2$ decay to heavy neutrinos, that is why these two channels make the "golden" process considered in the next section large. 
As the mixing in Eq.(\ref{mix}) becomes smaller, the $l \nu$ decay mode falls, e.g. for $M_D=0.1$ GeV we obtain $Br(W_2 \rightarrow l \nu)\simeq 10^{-11}$. 
These are a kind of textbook results, see e.g. \cite{mohbook} and references therein.

However, there are scenarios in which branching ratios can be different and heavy particles can decay dominantly to the light particles, so not through the right-handed currents. This is a case of non see-saw models where mixing angles are independent of heavy neutrino masses, see e.g. \cite{tev}.

Let us assume then that light-heavy neutrino mixing defined in Eq.(\ref{mix}) is independent of the heavy neutrino mass, experimental limits on elements of this mixing read (this limit has improved substantially over the last decade) \cite{delAguila:2008pw}

\begin{equation}
\sum_{j=4,5,6} U_{\nu_1, {j-3}} U_{ \nu_1, j-3}^{\ast} = U_{\nu_1, 4} U_{\nu_1, 4}^{\ast} \leq 0.003 \equiv \kappa^2_{max}.
\label{mixmax}
\end{equation}
\\  
In this case, the $l \nu$ branching ratio in Fig.\ref{totW2} will enhance\footnote{In a case where more than one heavy neutrino state exists (which is true in MLRSM), the maximal light-heavy neutrino mixing
defined in Eq.(\ref{mixmax}) is constrained further among others by neutrinoless double beta decay measurements 
to be less than $\kappa^2_{max}/2$ \cite{hep-ph/9703215}. 
We take then this parameter in our considerations for non-decoupling light-heavy neutrino mixings.}, BR($W_2 \to l \nu = 5\cdot 10^{-4}$). Still, it is not large.
$q \bar{q}$ and $l N$ modes dominate.

\FIGURE{
{\includegraphics[width=.49\textwidth]{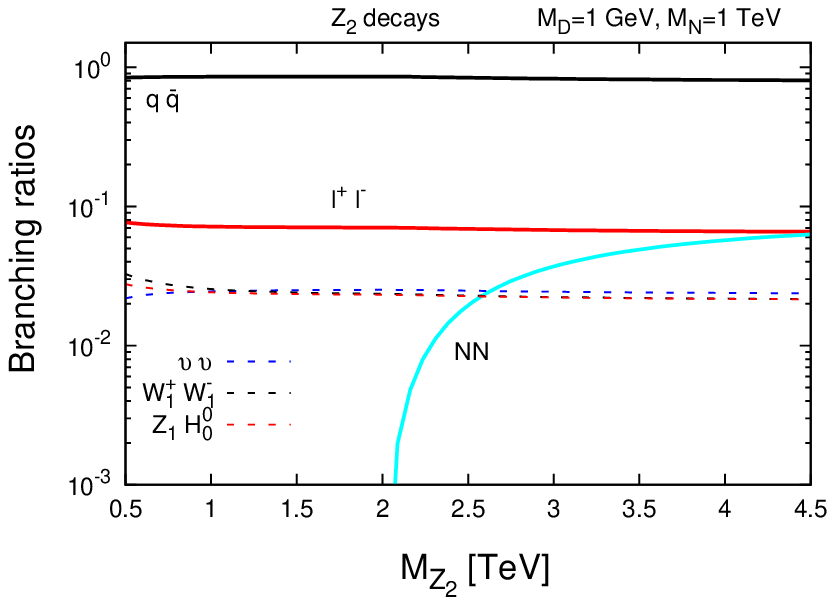}}
{\includegraphics[width=.49\textwidth]{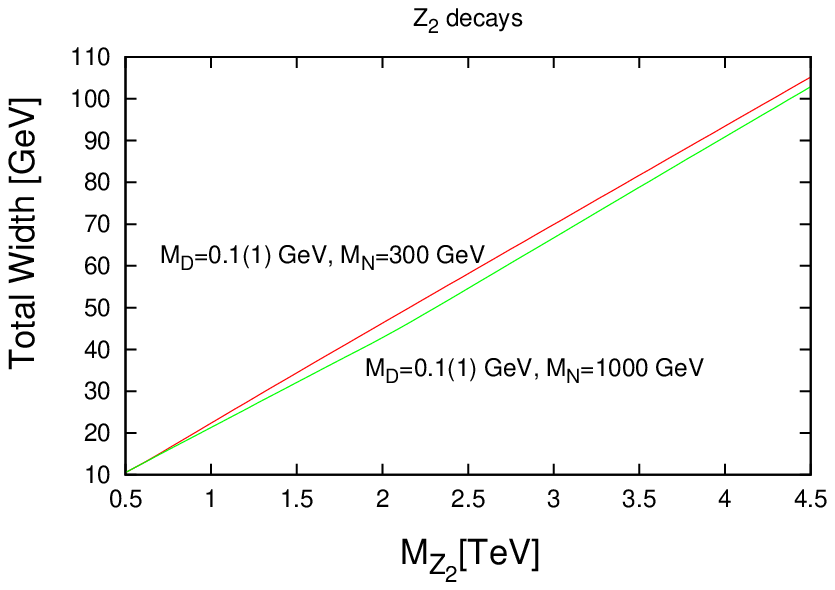}}
\caption{{Decay branching fractions  and total widths for $Z_2$ decays.}} 
\label{decayZ2}
}

\FIGURE{
{\includegraphics[width=.49\textwidth]{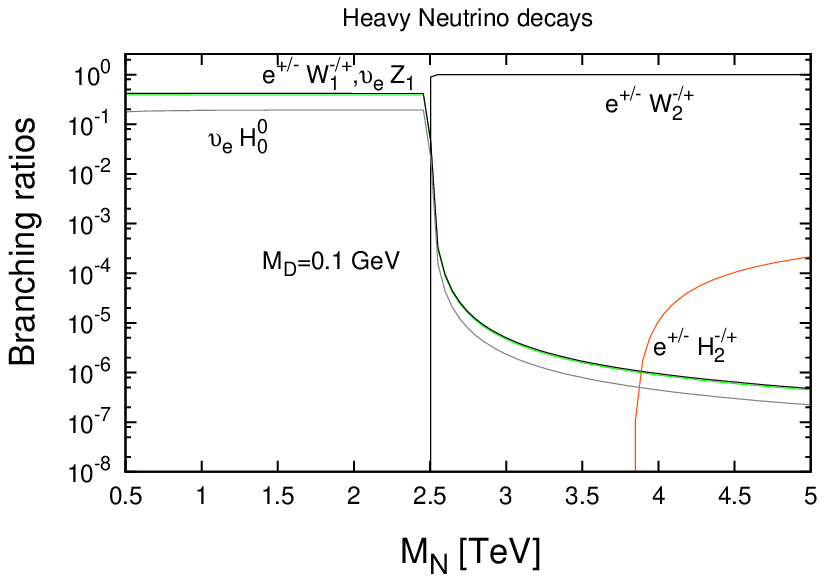}}
{\includegraphics[width=.49\textwidth]{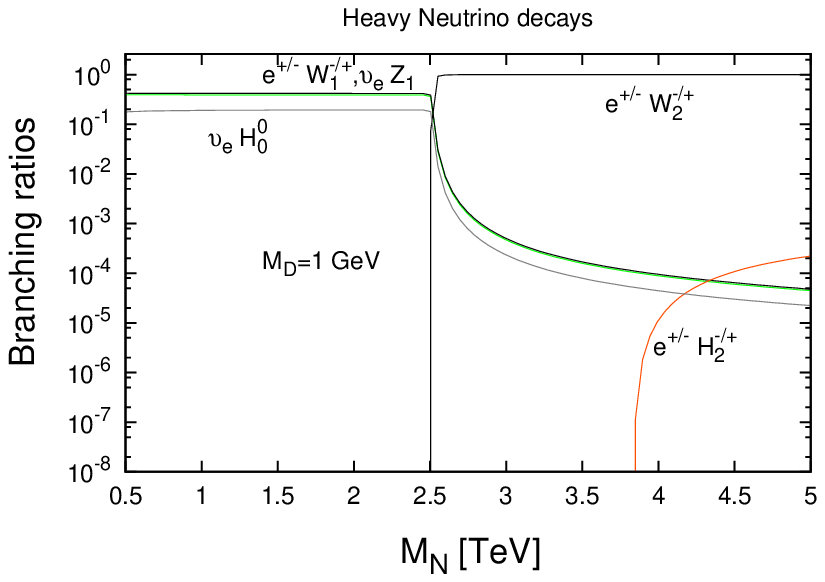}}
{\includegraphics[width=.49\textwidth]{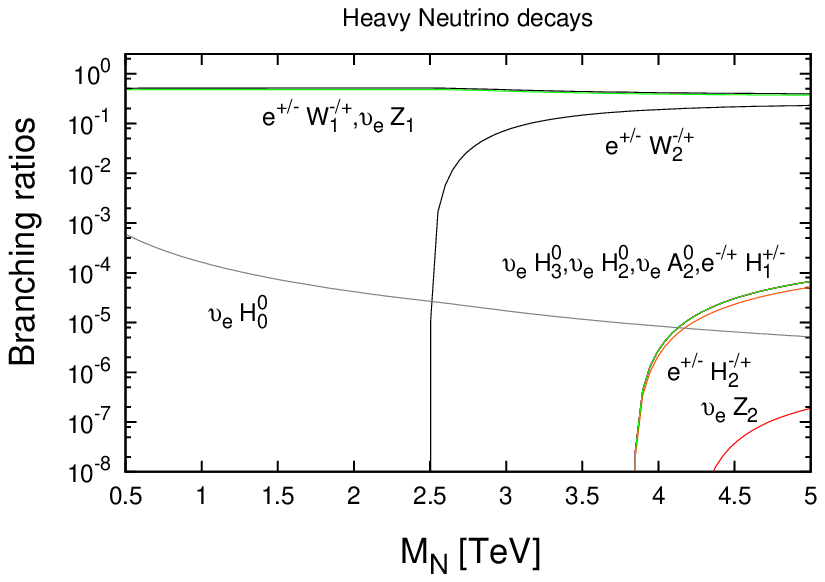}}
{\includegraphics[width=.49\textwidth]{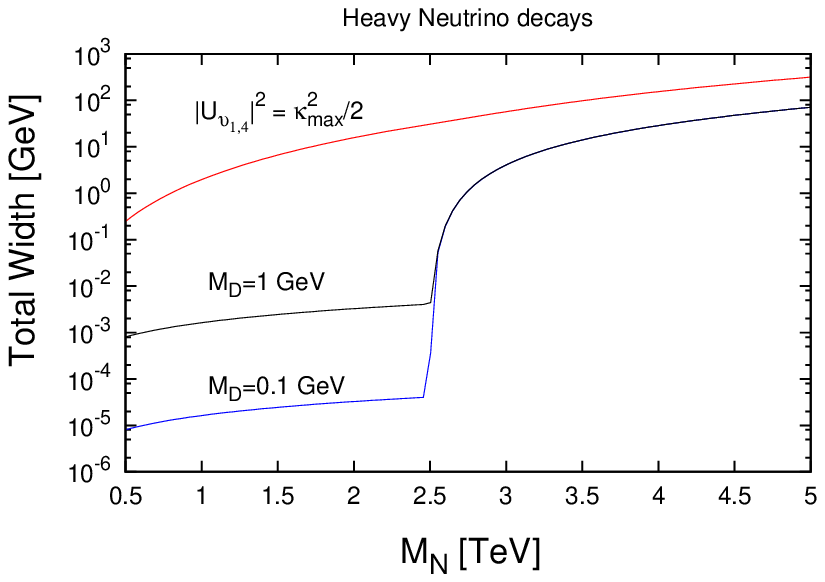}}
\caption[]{{Decay branching fractions and total widths for heavy neutrino decays with see-saw type of mixing, Eq.(\ref{mix}), first row
($\langle M_D \rangle$ = 0.1 GeV (on left) and 1 GeV (on right)). In the second row, on left, branching ratios with maximal type of mixings are calculated, Eq.(\ref{mixmax}). On right the total widths are given. $M_{W_2}$ is fixed at 2.5 TeV.
}} 
\label{decrn12}
}

In Fig.\ref{decayZ2} decays of the $Z_2$ boson are shown. Also here results are practically independent of light-heavy mixing scenarios, 
Eqs.(\ref{mix},\ref{mixmax}). 
$Z_2$ heavy boson decays are also dominated by quark channels. Here the situation is more complicated and up to a per mil level, a few channels contribute. Interestingly, also $Z_2$ decay to a pair of light neutrinos or bosons as well as to the $Z_1 H_0^0$ pair are substantial.

However, the situation changes with respect to light-heavy mixing scenarios for the case of heavy neutrino decays.
Decays of the first of heavy neutrinos $N_4$ in Fig.\ref{decrn12} are dominated by the $e^{\pm} W_1^{\mp}$ (we neglect here the mixing between different generations) mode till the threshold where $W_2$ production 
is open. Mass of $W_2$ is fixed at 2.5 TeV.  Still $e^{\pm} W_1^{\mp}$ option is large, even if $W_2$ mass would be smaller ($1.5\; {\rm TeV} \leq M_{W_2}
\leq 2.5$ TeV). Changing the mixing Eq.(\ref{mix}) affects mainly $\nu_e Z_2$ mode (which is negligible).

If we took maximal possible
mixing, $\kappa^2_{max}/2=0.0015$, then the branching ratios for heavy neutrino decays change qualitatively
(left figure in the second row in Fig.\ref{decrn12}).
We can see that the $N_4 \to e W_1$ and $N_4 \to \nu Z_1$ decays dominate over decay channels to the heavy states $W_2,Z_2$ in the 
kinematically allowed regions. The reason is that although decay amplitudes for light boson modes are proportional to the small light-heavy neutrino
mixing, the helicity summed amplitudes for gauge boson modes are suppressed in addition by the masses of gauge bosons, which is a stronger effect in a case of heavy gauge bosons. 
The difference between both scenarios of neutrino mixings is clearly visible on the last plot in Fig.\ref{decrn12} where total decay widths are given.

In order to show influence of the Higgs sector we deliberately distorted  heavy Higgs mass spectrum to include some lighter Higgs masses
such that Higgs particles show up in the neutrino decay. However, open in this way Higgs decay modes contribute well below per mille
level in total and are negligible. 
  
\subsection{LR signals at LHC, a sample}

The so-called "golden" process  where the left-right symmetry signal is not 
suppressed due to small light-heavy neutrino mixings  is depicted in Fig.\ref{udlljj} (here heavy neutrino couples directly to $W_2$ which decays hadronically, Fig.\ref{totW2}). 
Thus the final state consists of the same sign di-leptons and jets which also carries a clear signature of lepton number violation. 
Even if we consider the leptonic decay modes of $W_2$, we can have 3-leptons and $missing~~ energy$ as our signal events.
The presence of one missing energy source allows to reconstruct $W_2$ fully, and then the reconstruction of the right-handed 
neutrino, $N_{i}$, helps to reduce the combinatorial backgrounds for this process.  
In \cite{Maiezza:2010ic} it has been discussed that the dominant background for this process is coming  
from $t \bar{t}$ events and is negligible beyond the TeV scale. In the other case where $W_2$ decays hadronically 
with the largest branching fraction, the invariant mass of the hardest jets plus one(two)
lepton(s) also allows to  reconstruct in a clean way the heavy neutrino $N$ and $W_2$ masses.

\FIGURE{
{\includegraphics[width=.6\textwidth]{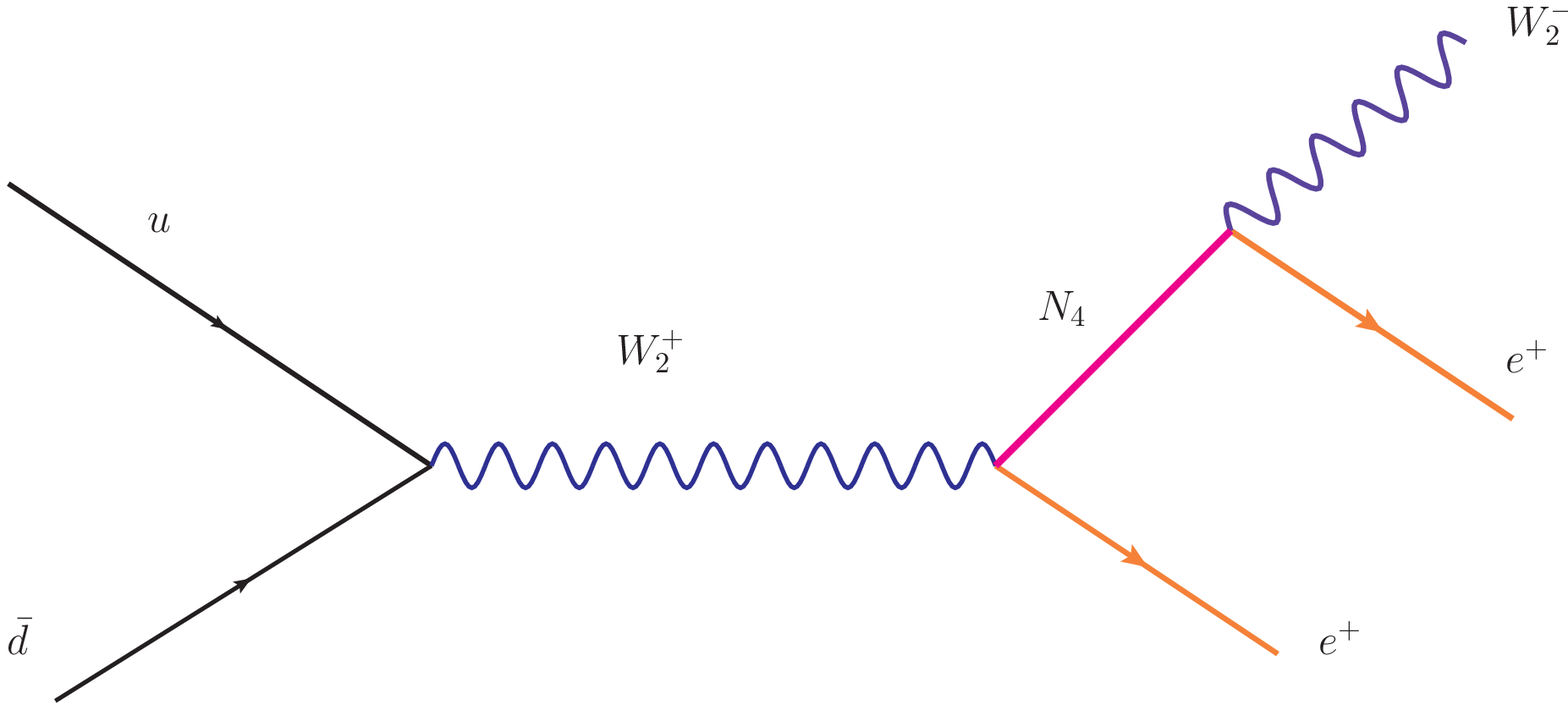}}
\caption[]{A tree level basic diagram for the $u\bar{d} \to e^{+}e^{+} W_2^-$ process. 
The process is not suppressed if $W_2^+$ decays to right-handed quarks forming jets 
(it is suppressed if it decays to standard, left-handed leptons).}
\label{udlljj}
}

\FIGURE{
{\includegraphics[angle=-90,width=.6\textwidth]{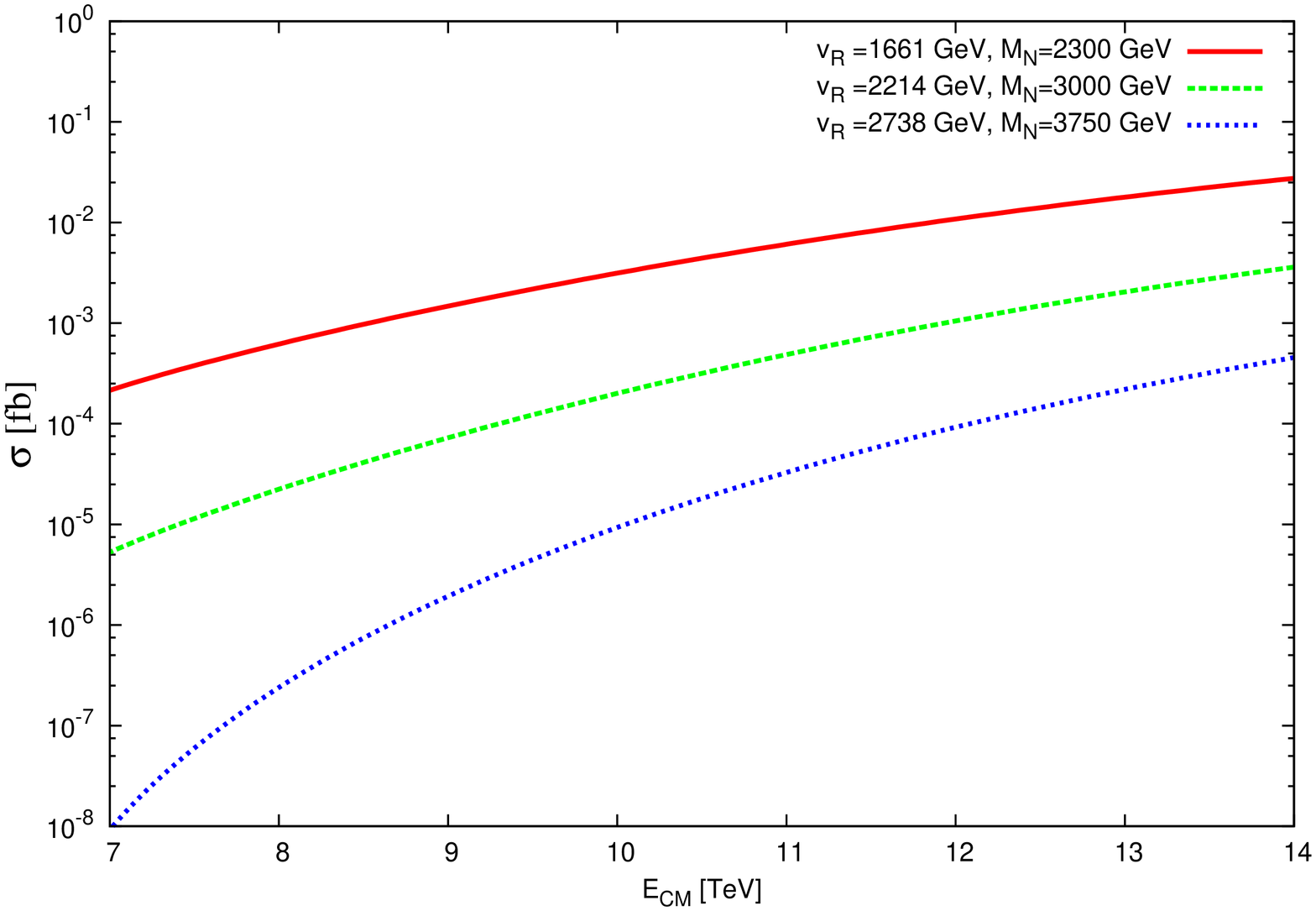}}
\caption{{Cross-sections for processes $pp\rightarrow e^{\pm} N_4 \rightarrow e ^{\pm} e^{\pm} W_2^{\mp}$ for sets
of parameters in Set B, $m_H=3,4,5$ TeV, respectively.}} 
\label{xsect-ern1-eew2}
}

As discussed in the last Section, muon decay data restricts very much possible values of ${\rm v}_R$ 
(and through the relation Eq.(\ref{mw2mz2}) masses of heavy gauge bosons) for chosen spectrum of Higgs and neutrino masses. 
Let us then assume a  scenario for LHC potential discoveries with $M_{W_2} \simeq 2.5$ TeV (then ${\rm v}_R \simeq 5$ TeV).
If we choose the most uniform scenario defined by {\bf Set B} in Table~\ref{tabrange1} (with the same masses for all 
Higgs particles and  also for all heavy neutrinos), then muon decay data sets the heavy neutrino masses of the order 
7 TeV (and masses of Higgs particles of the order of 10 TeV). 

We have computed the cross-section\footnote{For numerical results we use CalcHEP \cite{calchep} and Madgraph5 \cite{mdgr} with 
our own implementation of the MLRSM model 
in Feynrules \cite{fr}. We made a couple of cross checks for correctness of implementations for neutrino and 
gauge boson mixings. Results for $e^-e^+ \to \nu N$ \cite{gl92}, $e^-e^- \to W^-_1 W^-_1$ \cite{eeww}, 
$e^- \gamma \to N W_{1}^-$ \cite{hep-ph/9703215}
and $pp \to l W_2$ \cite{Ferrari:2000sp} have been recovered, among others.} for the process $pp\rightarrow e^{\pm} N_4 \rightarrow e ^{\pm} e^{\pm} W_2^{\mp}$ 
for the sets of parameters given in Set B (Table 1, $v_R=1661$ GeV is for $m_H=3$ TeV). 
Signatures for heavy neutrinos and charged gauge bosons in hadron colliders have been discussed already some decades ago, for a first paper on these kind of signals, see \cite{Keung:1983uu}.
Results are shown in Fig.\ref{xsect-ern1-eew2}. As can be seen in this case the cross-section is very small for LHC operating at 7 TeV, results for higher $m_H$ will give even smaller values. Going to 14 TeV of course improve the situation but still this scenario is very unlikely to be discovered.    

\FIGURE{ 
{\includegraphics[angle=-90,width=.6\textwidth]{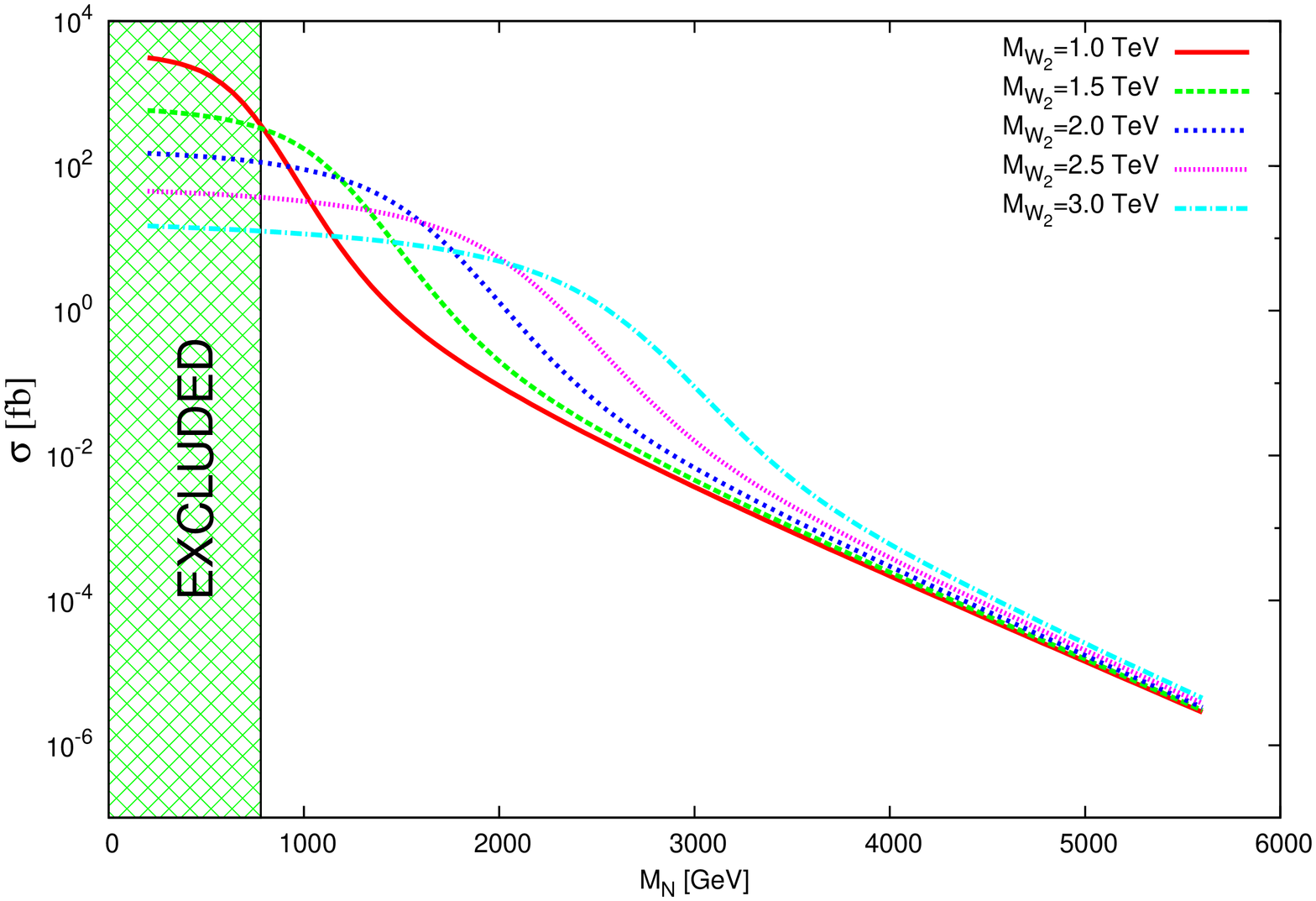}}
\caption[]{Cross-section for process $pp\longrightarrow e^{+} N_4$ as a function of heavy neutrino mass $M_N$ for different sets of $W_2$ masses, $\sqrt{s}=14$ TeV.
Excluded region of $M_N$ depends on $M_{W_2}$, see the plots in \cite{atlas, atlaslast}. We have just fixed it safely at $M_N=780$ GeV.}
\label{xsect-ern1a}
}

Luckily, other scenarios are possible  where one of heavy right-handed neutrinos has smaller mass, e.g. 800 GeV, but other two are very heavy having masses $\sim$ 5 TeV. There is also an option with 3 degenerate heavy neutrinos but with smaller ${\rm v}_R$, e.g. ${\rm v}_R=6398$ GeV ($M_{W_2}
\simeq 3$ TeV), see the last column in Table~\ref{tabrange1}.  
These scenarios are still compatible with muon decay data (though relatively light heavy gauge boson is required).
It gives much bigger cross-section,  see Fig.\ref{xsect-ern1a}, with anticipated luminosity 
this is a detectable process. 

From the above plot it is clear that as the mass of the heavy neutrino and the scale ${\rm v}_R$ increase, the production 
cross-section falls rapidly and then the further decays of the $N$ followed by the decay of $W_2$ suppress the effective cross-section 
for this "golden" process.

\FIGURE{
{\includegraphics[angle=-90,width=.6\textwidth]{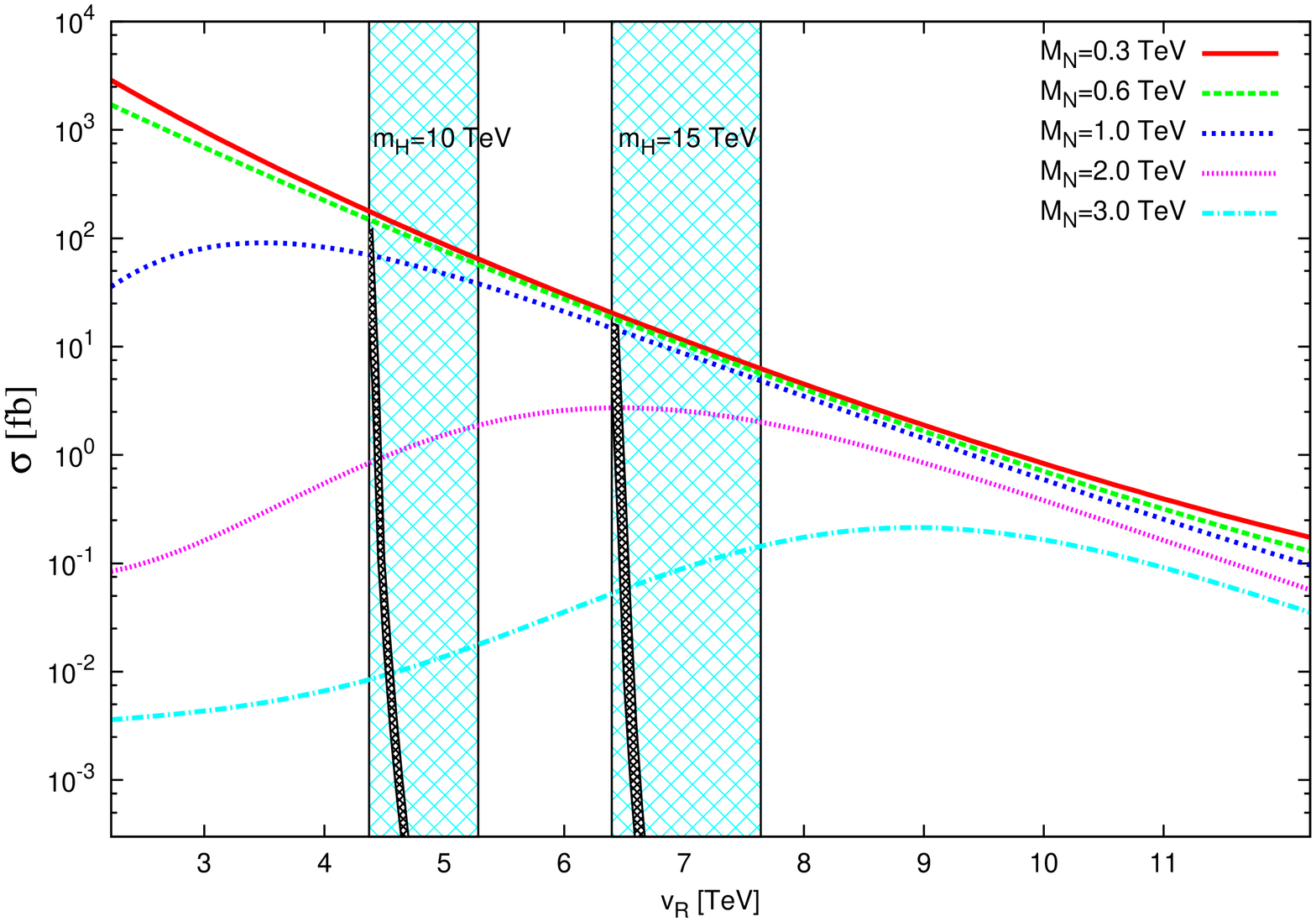}}
\caption[]{{Cross-section for process $pp\longrightarrow e^{+} N_4$ for different sets of heavy neutrino masses $M_N$, $\sqrt{s}=14$ TeV.
The results are for degenerate heavy neutrino and Higgs particle masses. $m_H$ masses are fixed at 10 TeV and $m_H=15$ TeV.
The whole shaded bands  correspond to parameters labeled as {\bf Set A} in Table~\ref{tabrange1} and Fig.\ref{scheme}. 
In addition, for each ${\rm v}_R$ between $({\rm v}_R)_{min}$ and
$({\rm v}_R)_B$ in Fig.~\ref{scheme}, heavy neutrino mass spectrum which is in agreement with muon decay data is obtained (we call it {\bf Set B}). 
In this way a possible cross-section for allowed $M_{W_2}-M_N$ masses is constrained dramatically. These regions are denoted by almost vertical and thin black stripes within the wider shaded regions.}} 
\label{xsect-ern1}
}
However, in Fig.\ref{xsect-ern1} we show more carefully how precise low energy data from Table~\ref{tabrange1} restricts a space of possible cross-section for this process.
Let us assume that Higgs masses are degenerate, at the level of 10 TeV and 15 TeV (the first case is almost excluded, see Eqs.(\ref{cms1}),(\ref{hm})
and Table~1). Then vertical bands restrict regions of possible 
cross-sections for given $M_N$ masses. If we assume in addition that heavy neutrino masses are also degenerate, then the black, thin strips inside these bands give for each ${\rm v}_R$ very narrow intervals of possible heavy neutrino masses, consequently, region of possible cross-sections is very limited.
With an assumed luminosity of tens of inverse femtobarns at $\sqrt{s}=14$ TeV,  $\sigma(pp  \to e^+ N)\simeq 10$ fb would give hundreds of events, which we take as a safety discovery limit for this process. 
Relevant experimental conditions do not spoil signals, for a discussion on kinematical cuts and a background for this process, see e.g. \cite{Ferrari:2000sp}. In this case, without muon data, 
possible ${\rm v}_R$ values give  $W_2$ mass in the range (1 {\rm TeV}, 3.5 {\rm TeV}) for heavy neutrino masses up to 1 TeV.
Muon data shrinks the region very much, $1970\; {\rm GeV} \leq M_{W_2} \leq 2050\; {\rm GeV}$ (for $m_H=10$ TeV) and   
$3008\; {\rm GeV} \leq M_{W_2} \leq 3040\; {\rm GeV}$ (for $m_H=15$ TeV). From Fig.\ref{xsect-ern1} it should be clear that increasing 
heavy Higgs masses would shift the $v_R$ scale to higher level, decreasing further cross sections for the considered process.   
In summary, for the left-right LHC phenomenology, Higgs mass spectrum is optimal in vicinity of 15 TeV region. 
For a case of degenerate heavy neutrinos, heavy Higgs particles with masses at about 10 TeV and below are practically excluded by muon data. 
On the other hand, MLRSM scenarios with Higgs particles masses at about 20 TeV (and above) are allowed by muon data, however,  
low energy muon restrictions constraint heavy gauge boson and neutrino masses in such a way that $\sigma(pp  \to e^+ N)<10$ fb.

\section{Conclusions}

It is very important to take into account low energy data in phenomenological analysis of non-standard models at LHC.
This is a quite common action in supersymmetric models, e.g. precise $(g-2)_\mu$ analysis is very important for pinning down parameter space  for
supersymmetry collider searches \cite{jegrep,olive}. These kind of analysis are less popular in GUT models (it is justified if a decoupling of heavy states occurs). We should acknowledge the last work \cite{Tello:2010am} where a connection between neutrinoless doubly beta decay and LHC for LR models is undertaken  (this is however by its nature purely "tree level" calculation and connection).

Here we show the interplay between fermion-boson heavy spectrum of the MLRSM model in the muon decay. As it is  typical for GUT models, it is also true for MLRSM that 
\emph{"extensions of the SM in most cases end up in a fine tuning problem, because decoupling of new
heavy states, in theories where masses are generated by spontaneous symmetry
breaking, is more the exception than the rule"} (quotation from \cite{fred}). 
As shown in Section \ref{seclow},
fixing heavy gauge boson masses and Higgs particle masses, the region of possible heavy neutrino mass spectrum is restricted by the muon decay. However, 
there is still a way to get at least one relatively light heavy neutrino, which can be explored at LHC. This is possible as heavy particles effects are effectively "weighted" at the 1-loop level (for virtual particles the effects are summed up which means that effects of 3 heavy degenerate neutrinos can be equivalent to the effects of one relatively light and two heavier heavy neutrino masses).
In this case, there is still a way that left-right symmetry is broken at low enough energy scale such that LR models can be discovered directly at LHC (see Section \ref{lhcex}). 

Let us note, that the situation gets more interesting if LHC finds  heavy particles which appear in the spectrum of the LR model.
Then analysis could be reversed  -- the obtained physical parameters can be
helpful to further pin down  remaining parameters for a part of the spectrum which can not be directly constrained at LHC, through the low energy precise analysis like the muon decay. For instance,  knowledge of both the mass of the lightest of heavy neutrinos and of the scale ${\rm v}_R$ ($W_2$ boson mass reconstruction) will restrict 
masses of heavier neutrinos in $\Delta r$. This will be a great hint for searches of remaining particles since we would be able to predict where to look for them. 

We think that this kind of low-high energy analysis is important and should be further explored, for instance including 1-loop level calculations in MLRSM for lepton flavour violating processes. In general, when making numerical predictions for any model beyond the SM, as many as possible of low energy observables and precision LEP observables should be taken into account. 

\section*{Acknowledgements}
We would like to thank Henryk~Czy\.z, Fred~Jegerlehner, Miha Nemev$\v{s}$ek and Marek~Zra{\l}ek for useful discussions and comments.
Work supported by the Research Executive Agency (REA) of the European Union under the
Grant Agreement number PITN-GA-2010-264564 (LHCPhenoNet) and by the Polish Ministry of Science under grant No. N N202 064936.


\begin{thebibliography}{52}


\bibitem{Czakon:2002wm}
  M.~Czakon, J.~Gluza, J.~Hejczyk,
  Nucl.\ Phys.\  {\bf B642 } (2002)  157-172.


\bibitem{npb} M. Czakon, J. Gluza,M. Zra\l ek,
  Nucl. Phys. {\bf B573} (2000) 57.


\bibitem{epj} M. Czakon, J. Gluza, F. Jegerlehner, M. Zra\l ek,
  Eur. J. Phys. {\bf C13} (2000) 275.


\bibitem{Chen:2005jx}
  M.-C.~Chen, S.~Dawson, T.~Krupovnickas,
  Int.\ J.\ Mod.\ Phys.\  {\bf A21 } (2006)  4045-4070.


\bibitem{Chankowski:2006hs}
  P.~H.~Chankowski, S.~Pokorski, J.~Wagner,
  Eur.\ Phys.\ J.\  {\bf C50 } (2007)  919-933.

\bibitem{Chankowski:2006jk}
  P.~H.~Chankowski, S.~Pokorski, J.~Wagner,
  Eur.\ Phys.\ J.\  {\bf C47 } (2006)  187-205.


  
\bibitem{pati2} J.C. Pati and A. Salam, Phys. Rev. {\bf D10} (1974) 275;
  R.N. Mohapatra, J.C.Pati, Phys. Rev. {\bf D11} (1975) 2558;
  R.N. Mohapatra, J.C.Pati, ibid. {\bf D11} (1975) 566;
  G. Senjanovic and R.N Mohapatra, ibid. {\bf D12} (1975) 1502;
  R.N. Mohapatra, P.B. Pal, Phys. Rev. {\bf D38} (1998) 2226;
  G. Senjanovic, Nucl. Phys. {\bf B153} (1979) 334.

\bibitem{georgi} H. Georgi and S.L. Glashow,
  Phys. Rev. Lett. {\bf 32} (1974) 438.  

\bibitem{class} J.~F.~Gunion, J.~Grifols, A.~Mendez, B.~Kayser and
  F.~I.~Olness, Phys.\ Rev.\ D {\bf 40} (1989) 1546;
  N.~G.~Deshpande, J.~F.~Gunion, B.~Kayser and F.~I.~Olness,
  Phys.\ Rev.\ D {\bf 44} (1991) 837.  


 
\bibitem{tev}
  J.~Gluza,
  Acta Phys.\ Polon.\  {\bf B33 } (2002)  1735-1746.




\bibitem{gl92} J. Gluza, M. Zralek, Phys. Rev. {\bf D48} (1993) 5093;
 Phys.\ Rev.\ D {\bf 51} (1995) 4695.

\bibitem{ann} P. Duka, J. Gluza and M. Zra\l ek,
  Ann. of Phys. {\bf 280} (2000) 336.
  



\bibitem{barbsn1987}
  R.~Barbieri and R.~N.~Mohapatra,
  Phys.\ Rev.\ D\ {\bf 39} (1989) 1229.

\bibitem{lang}
  P.~Langacker and S.~Uma Sankar,
  Phys.\ Rev.\ D\ {\bf 40} (1989) 1569.


  
\bibitem{pdg2011} K. Nakamura et al. (Particle Data Group), J. Phys. G 37, 075021 (2010). 

 

\bibitem{cms}
  S.~Chatrchyan {\it et al.} [CMS Collaboration],
  Phys.\ Lett.\ B\ {\bf 701} (2011) 160.

\bibitem{cms2}
The CMS collaboration, CMS preprint PAS EXO-11-002.


\bibitem{atlas}
The ATLAS collaboration, "A Search for Heavy Majorana Neutrino and $W_R$ in
dilepton plus jets events with the ATLAS detector in pp collisions at $\sqrt{s} = 7$ TeV",  
ATLAS-CONF-2011-115.

\bibitem{atlaslast}
  G.~Aad {\it et al.}  [ATLAS Collaboration],
  arXiv:1203.5420 [hep-ex].
\bibitem{klks1} Y.~Zhang, H.~An, X.~Ji and R.N.~Mohapatra, Nucl. Phys. {\bf B802}, 247 (2008).

\bibitem{Maiezza:2010ic}
  A.~Maiezza, M.~Nemevsek, F.~Nesti, G.~Senjanovic,
  Phys.\ Rev.\  {\bf D82 } (2010)  055022.
  

\bibitem{Chakrabortty:2012mh}
  J.~Chakrabortty, H.~Z.~Devi, S.~Goswami and S.~Patra,
  arXiv:1204.2527 [hep-ph].

 



\bibitem{plb} M.~Czakon, J.~Gluza and M.~Zralek,
  Phys.\ Lett.\ B {\bf 458} (1999) 355. 
  
 \bibitem{Nemevsek:2011hz}
  M.~Nemevsek, F.~Nesti, G.~Senjanovic, Y.~Zhang,
  Phys.\ Rev.\  {\bf D83 } (2011)  115014.



\bibitem{Bansal:2009jx}
  V.~Bansal,
  arXiv:0910.2215 [hep-ex].


 
\bibitem{Ferrari:2000sp}
  A.~Ferrari {\it et al.},
  Phys.\ Rev.\  {\bf D62 } (2000)  013001.




  \bibitem{Shaban:1992vv}
  N.~T.~Shaban and W.~J.~Stirling,
  Phys.\ Lett.\ B {\bf 291} (1992) 281.
  
  
\bibitem{Lindner:1996tf}
  M.~Lindner and M.~Weiser,
  Phys.\ Lett.\ B {\bf 383} (1996) 405.

\bibitem{beal} G.~Beall, M.~Bander and A.~Soni,
Phys.\ Rev.\ Lett.\  {\bf 48} (1982) 848.
  
\bibitem{pil} 
A.~Pilaftsis,
Phys.\ Rev.\ D {\bf 52} (1995) 459;
Z.~Gagyi-Palffy, A.~Pilaftsis and K.~Schilcher,
Nucl.\ Phys.\ B {\bf 513} (1998) 517.


\bibitem{bsg}  
J.~M.~Frere and J.~Matias,
  Nucl.\ Phys.\ B {\bf 572} (2000) 3;
  M.~E.~Pospelov, Phys.\ Rev.\ D {\bf 56} (1997) 259;
  K.~Kiers, et al., Phys.\ Rev.\ D {\bf 66} (2002) 095002;
  T.G. Rizzo, Phys. Rev. {\bf D50} (1994) 3303;
  P.~Cho, M. Misiak, Phys. Rev. {\bf D49} (1994) 5894.

\bibitem{sok} 
G.~Senjanovic and A.~Sokorac,
Phys.\ Lett.\ B {\bf 76} (1978) 610;
G.~Senjanovic and A.~Sokorac,
Phys.\ Rev.\ D {\bf 18} (1978) 2708.


 
 \bibitem{Tello:2010am}
  V.~Tello, M.~Nemevsek, F.~Nesti, G.~Senjanovic, F.~Vissani,
  Phys.\ Rev.\ Lett.\  {\bf 106 } (2011)  151801.

\bibitem{Frank:2011rb}
  M.~Frank, A.~Hayreter, I.~Turan,
Phys.\ Rev.\ D {\bf 84} (2011) 114007; Phys.Rev. D83 (2011) 035001.

\bibitem{Jezo:2012rm}
  T.~Jezo, M.~Klasen and I.~Schienbein,
  arXiv:1203.5314 [hep-ph].
  
 \bibitem{matching}
 M.~Awramik and M.~Czakon,
  Phys.\ Lett.\ B {\bf 568} (2003) 48.
  
\bibitem{FCNC}
  D.~Guadagnoli and R.~N.~Mohapatra,
  Phys.\ Lett.\ B {\bf 694} (2011) 386.

\bibitem{eeww}
  J.~Gluza and M.~Zralek,
  Phys.\ Rev.\ D\ {\bf 52} (1995) 6238;
  J.~Gluza,
  Phys.\ Lett.\ B\ {\bf 403} (1997) 304.



\bibitem{triplegauge}
E.~Arik et al., "A study of $pp \to W' \to WZ$ at LHC in the ATLAS experiment",
ATL-PHYS-2001-005.
 
\bibitem{mohbook} R.N.Mohapatra, "Unification and Supersymmetry", Springer-Verlag, 2003.

\bibitem{hep-ph/9703215}
  J.~Gluza, J.~Maalampi, M.~Raidal and M.~Zralek,
  Phys.\ Lett.\ B\ {\bf 407} (1997) 45.

\bibitem{Keung:1983uu}
  W.~-Y.~Keung and G.~Senjanovic,
  Phys.\ Rev.\ Lett.\  {\bf 50} (1983) 1427.
  
\bibitem{delAguila:2008pw}
  F.~del Aguila, J.~de Blas and M.~Perez-Victoria,
  Phys.\ Rev.\ D {\bf 78} (2008) 013010.
  
\bibitem{calchep}
  A.~Pukhov,
  arXiv:hep-ph/0412191.


\bibitem{mdgr}
  J.~Alwall, M.~Herquet, F.~Maltoni, O.~Mattelaer and T.~Stelzer,
  JHEP\ {\bf 1106} (2011) 128.
  
  \bibitem{fr}
  N.~D.~Christensen and C.~Duhr,
  Comput.\ Phys.\ Commun.\ \ {\bf 180} (2009) 1614.
 
\bibitem{jegrep}
  F.~Jegerlehner and A.~Nyffeler,
  Phys.\ Rept.\ \ {\bf 477} (2009) 1.

\bibitem{olive}
  K.~A.~Olive,
  Eur.\ Phys.\ J.\ C\ {\bf 59} (2009) 269.
 
\bibitem{fred} 
F.~Jegerlehner,
  arXiv:1110.0869 [hep-ph].
 
\end{thebibliography}
\end{document}